\documentstyle[aps,prd,epsfig,amstex]{revtex}

\newcommand{\vv}[1]{{\mathbf{#1}}}
\newcommand{\uv}[1]{\hat{\mathbf{#1}}}
\newcommand{\op}[1]{{\mathbf{#1}}}

\begin{document}

\draft

\wideabs{
\title{Data analysis strategies for the detection of gravitational waves in
       non-Gaussian noise}
\author{Jolien D. E. Creighton}
\address{Theoretical Astrophysics,
         California Institute of Technology,
         Pasadena, California 91125}
\date{26 January 1999}
\maketitle
\begin{abstract}%
In order to analyze data produced by the kilometer-scale gravitational wave
detectors that will begin operation early next century, one needs to develop
robust statistical tools capable of extracting weak signals from the detector
noise.  This noise will likely have non-stationary and non-Gaussian
components.  To facilitate the construction of robust detection techniques, I
present a simple two-component noise model that consists of a background of
Gaussian noise as well as stochastic noise bursts.  The optimal detection
statistic obtained for such a noise model incorporates a natural veto which
suppresses spurious events that would be caused by the noise bursts.  When two
detectors are present, I show that the optimal statistic for the non-Gaussian
noise model can be approximated by a simple coincidence detection strategy.
For simulated detector noise containing noise bursts, I compare the operating
characteristics of (i) a locally optimal detection statistic (which has
nearly-optimal behavior for small signal amplitudes) for the non-Gaussian
noise model, (ii) a standard coincidence-style detection strategy, and
(iii) the optimal statistic for Gaussian noise.
\end{abstract}
}

\narrowtext
The reliable detection of weak gravitational wave signals from broad-band
detector noise from kilometer-scale interferometers such as
LIGO~\cite{abramovici:1992} and VIRGO~\cite{caron:1997} is the primary concern
in developing gravitational wave data analysis strategies.  Because of the
weakness of the expected gravitational wave signals, it is critical that the
detection strategy should be nearly optimal.  However, the detector noise may
not be purely stationary and Gaussian, so it is important that the detection
strategy also be robust so that detections will be reliable.

Until now, most work on the development of data analysis strategies has been
limited to the case of stationary Gaussian noise, though there has been some
work \cite{grasp} on the creation of vetoes that will discriminate between
expected gravitational wave signals and non-Gaussian noise bursts.  Because
the properties of the noise in the LIGO and VIRGO detectors will not be known
in advance, it is difficult to assess how well these strategies and vetoes
will perform.  When real interferometer data from the 40-meter Caltech
prototype is used, it is found that additional vetoes are needed to deal with
the abundance of false alarms arising from the non-Gaussian noise
\cite{allen:1998,creighton:1998}.

In this paper, I present a simple non-Gaussian noise model, consisting of
Poisson-distributed noise bursts, that represents a number of potential
non-Gaussian noise sources that may be present in future interferometers.
This noise model is less na\"{\i}ve than the usual assumptions of Gaussian
noise alone, but is simple enough that many analytical results can be
obtained.  By using this noise model, the robustness of various detection
strategies can be assessed.  A general introduction to signal detection in
non-Gaussian noise can be found in Ref.~\cite{kassam:1988}.  The new result in
this paper is the use of the non-Gaussian noise model in examining the
performance of simple multi-detector search strategies, which will be
important for gravitational wave searches.

I will adopt the following notation in this paper.  The detector output is a
set of $N$ values that are written collectively as a vector $\vv{h}$ in a
$N$-dimensional vector space~$V$.  This vector can be thought of as having
components $h_j=(\vv{h},\vv{e}_j)$, $j\in[0,N-1]$, which represent a time
series of sample measurements made by the detector.  Here, $\vv{e}_j$ is the
appropriate Cartesian basis on $V$.  Alternatively, the vector can be
expressed as the set of components $\{\tilde{h}_{2k},\tilde{h}_{2k+1}\}$,
$k\in[0,N/2-1]$---the real and imaginary Fourier transform components of
$\{h_j\}$.  These components are given by
$\tilde{h}_j=(\vv{h},\tilde{\vv{e}}_j)$, where
$\tilde{\vv{e}}_{2k}=(2/N)^{1/2}\sum_{j=0}^{N-1}\vv{e}_j\cos(2\pi jk/N)$ and
$\tilde{\vv{e}}_{2k+1}=(2/N)^{1/2}\sum_{j=0}^{N-1}\vv{e}_j\sin(2\pi jk/N)$,
$k\in[0,N/2-1]$.  Thus, the vectors can be treated in either a time
representation or a frequency representation.

A natural inner product of two vectors, $\vv{a}$ and $\vv{b}$, in this vector
space is defined as $(\vv{a},\op{Q}\vv{b})$.  The kernel $\op{Q}$ is the
inverse of the auto-correlation matrix, $\op{R}$, of the Gaussian component,
$\vv{n}_{\text{G}}$, of the detector noise.  Thus,
$\op{R}=\langle\vv{n}_{\text{G}}\otimes\vv{n}_{\text{G}}\rangle$, where the
angle brackets denote an average over an ensemble of realizations of detector
noise.  Vector norms are defined in terms of this inner product, i.e.,
$\|\vv{a}\|^2=(\vv{a},\op{Q}\vv{a})$, as are unit vectors:
$\uv{a}=\vv{a}/\|\vv{a}\|$.

The detector noise $\vv{n}$ consists of two components: (i) the Gaussian
component $\vv{n}_{\text{G}}$ (that is always present) and (ii) a possible
noise burst component $\vv{n}_{\text{B}}$ that is present with probability
$P_{\text{B}}$.  The Gaussian component has the probability distribution
$p[\vv{n}_{\text{G}}]\propto\exp(-\|\vv{n}_{\text{G}}\|^2/2)$.  If the burst
component is randomly distributed in the vector space $V$ with normalized
measure $\hat{D}[\vv{n}_{\text{B}}]$, then the probability distribution for
the noise is
\begin{equation}
  p[\vv{n}] \propto e^{-\|\vv{n}\|^2/2} + \frac{P_{\text{B}}}{1-P_{\text{B}}}
    \int \hat{D}[\vv{n}_{\text{B}}] e^{-\|\vv{n}-\vv{n}_{\text{B}}\|^2/2}.
\end{equation}
Suppose that the noise bursts are uniformly-distributed in the vector space
$V$ out to a large radius $R$.  Such bursts will typically last the entire
duration of interest ($N$ samples) and fill the entire frequency band.  The
noise distribution is approximately
\begin{equation} \label{e:noiseBurst}
  p[\vv{n}] \propto e^{-\|\vv{n}\|^2/2}
    ( 1 + \epsilon e^{\|\vv{n}\|^2/2} )
\end{equation}
for $\|\vv{n}\|<R$, where
$\epsilon\simeq2^{N/2}\Gamma(N/2+1)R^{-N}P_{\text{B}}/(1-P_{\text{B}})$ for
large $R$.

An alternative derivation of Eq.~(\ref{e:noiseBurst}) is the following.
Suppose the noise burst is simply an additional transient source of Gaussian
noise.  Then Eq.~(\ref{e:noiseBurst}) can be written as
\begin{equation} \label{e:GaussianNoiseBurst}
  p[\vv{n}] \propto e^{-(\vv{n},\op{Q}\vv{n})/2}
    (1 + \varepsilon e^{(\vv{n},[\op{Q}-\op{Q}']\vv{n})/2}),
\end{equation}
where $\varepsilon=(\det|\op{R}|/\det|\op{R}'|)^{N/2}
P_{\text{B}}/(1-P_{\text{B}})$, $\op{R}'$ is the auto-correlation matrix for
both ambient and the burst components of Gaussian noise together, and the
matrix $\op{Q}'$ is the inverse of $\op{R}'$.  If the typical noise burst is
much louder than the ambient Gaussian noise, then
$\op{Q}-\op{Q}'\approx\op{Q}$.  Therefore, in the case of loud Gaussian noise
bursts, Eq.~(\ref{e:GaussianNoiseBurst}) has the same form as
Eq.~(\ref{e:noiseBurst}).

The likelihood ratio can now be computed for the two alternative hypotheses:
that the output $\vv{h}=\vv{n}$ is noise alone ($H_0$), or that the output
$\vv{h}=A\uv{u}+\vv{n}$ is a signal $A\uv{u}$ of amplitude $A$ embedded in
noise ($H_1$).%
\footnote{In this paper, I consider only signals which are completely known
(up to their amplitude).  The generalization of this case to one in which the
signals have an unknown initial phase is straightforward: see, e.g.,
Ref.~\cite{wainstein:1962}.}
The likelihood ratio $\Lambda(A)=p[\vv{h}\mid H_1]/p[\vv{h}\mid H_0]$ is the
ratio of the posterior probability of obtaining the observed output $\vv{h}$
given hypothesis $H_1$ to the posterior probability of obtaining $\vv{h}$
given hypothesis $H_0$.  One then finds that the likelihood ratio is
\begin{equation} \label{e:likelihoodRatioA}
  \Lambda(A) = \frac{\Lambda_{\text{G}}(A) + \alpha}{1 + \alpha}
\end{equation}
where
\begin{equation} \label{e:likelihoodRatioG}
  \Lambda_{\text{G}}(A) = e^{A(\uv{u},\op{Q}\vv{h}) - A^2/2}
\end{equation}
and
\begin{equation}
  \alpha = \epsilon e^{\|\vv{h}\|^2/2}.
\end{equation}
The quantity $\Lambda_{\text{G}}(A)$ is the likelihood ratio one would obtain
if only the Gaussian noise component were present; it is a monotonically
increasing function of the matched filter $(\uv{u},\op{Q}\vv{h})$.  When a
non-Gaussian noise burst component can occur, the likelihood ratio depends on
an additional measured quantity: the magnitude of the output vector,
$\|\vv{h}\|$.  In addition to these functions of the detector output, the
likelihood ratio also depends on the expected amplitude $A$ of the signal and
the proportionality factor $\epsilon$, which encapsulates the probability of a
noise burst and the maximum possible amplitude $R$ of the burst.

The quantity $\alpha$ plays a central role in the modified likelihood ratio of
Eq.~(\ref{e:likelihoodRatioA}): it acts as a detector of noise bursts, and
vetoes events that are more likely due to the burst.  The logarithm of
$\alpha$ is proportional to the amount of power in the detector output.  For
Gaussian noise alone, the expected value of the power is
$\langle\|\vv{n}_{\text{G}}\|^2\rangle=N$, and thus
$\langle\ln\alpha\rangle\sim\ln P_{\text{B}}-\case{1}{2}N\ln(R^2/N)$ for large
$N$.  Thus, for $R^2>N$ (as was assumed above), the value of $\alpha$ will be
small.  However, when a noise burst is present, the value of $\ln\alpha$ is
increased by a typical amount $\sim\case{1}{2}R^2$, so $\alpha$ will typically
become large.  In fact, $\alpha$ is the excess-power statistic for detection
of arbitrary noise bursts \cite{flanagan:1998,anderson:1998}.  For small
values of $\alpha$, the likelihood ratio approaches the usual Gaussian noise
likelihood ratio.  However, for large values of $\alpha$, the likelihood ratio
approaches unity.

\begin{figure}[ht]
\epsfig{file=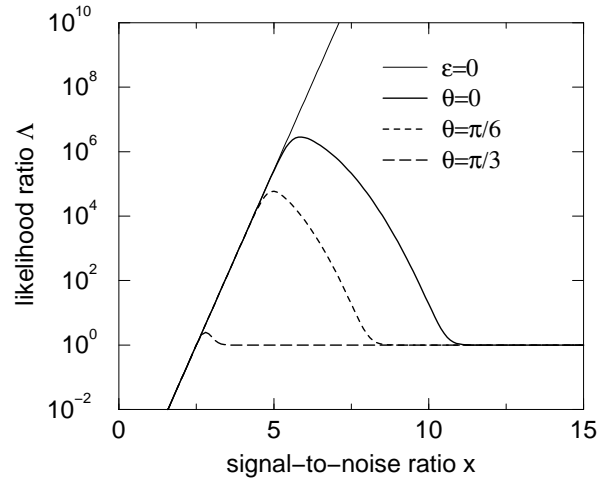,width=0.9\linewidth}
\caption{Likelihood ratio as a function of the matched filter signal-to-noise
ratio $x=(\uv{u},\op{Q}\vv{h})$ and the direction cosine
$\cos\theta=(\uv{u},\op{Q}\uv{h})$.  The model signal and noise burst both lie
in an $N=4$ dimensional vector space $V$.  The model signal has an
amplitude $A=5$.  Noise bursts occur with a probability $P_{\text{B}}=1\%$; the
possible burst vectors are uniformly distributed in $V$ out to a maximum
radius~$R=25$.  The thin solid line is the likelihood ratio if no noise
bursts were present ($\epsilon=0$), while the thick dotted lines are plots of
Eq.~(\ref{e:likelihoodRatio}) with $\epsilon=2\times10^{-7}$ (corresponding to
the above parameters).
\label{f:logLambda}}
\end{figure}

The likelihood ratio is a function of the detector output via the two
quantities $x=(\uv{u},\op{Q}\vv{h})=\|\vv{h}\|\cos\theta$ and $\|\vv{h}\|$.
The likelihood ratio also depends on the expected signal amplitude $A$ and the
factor $\epsilon$.  Thus, the likelihood ratio is
\begin{equation} \label{e:likelihoodRatio}
  \Lambda(A) = \frac{e^{Ax-A^2/2} + \epsilon e^{(x\sec\theta)^2/2}}
    {1 + \epsilon e^{(x\sec\theta)^2/2}}.
\end{equation}
Figure~\ref{f:logLambda} shows the likelihood ratios as functions of the
matched filter statistic $x$ and the angle $\theta$ for a given value of
$\epsilon$ and $A$.  Notice that the likelihood ratio is attenuated when the
magnitude of the output, $\vv{h}$, is much larger than the largest expected
signal; this attenuation occurs at smaller signal-to-noise ratios for smaller
absolute values of the direction cosine $\cos\theta$.

Because of the non-trivial dependence of the likelihood ratio on the expected
signal amplitude, the optimal statistic for signals of some amplitude $A$ will
not be the same statistic as the optimal statistic for a different amplitude
$A'$.  This situation is unlike the case of purely Gaussian noise, for which
the likelihood ratio grows monotonically with the matched filter output,
regardless of the expected signal amplitude.  Because the amplitude of a
signal will not typically be known in advance, it is useful to consider a
\emph{locally optimal} statistic \cite{kassam:1988}, which provides the
optimal performance in the limit of small amplitude signals.  The rationale
behind such a choice is that large amplitude signals pose no challenge for
detection, so any reasonable statistic will suffice for them.  The locally
optimal statistic, which is defined as
$\lambda=d\ln\Lambda/dA|_{A=0}=(\uv{u},-\bbox{\nabla}\ln p[\vv{h}])$, is given
by
\begin{equation} \label{e:localOpt}
  \lambda = \frac{x}{1+\epsilon e^{(x\sec\theta)^2/2}},
\end{equation}
where $x=(\uv{u},\op{Q}\vv{h})$ is the matched filter.  Notice that the
locally optimal statistic also incorporates a veto based on the value of
$\alpha=\epsilon\exp[(x\sec\theta)^2/2]$.  The locally optimal statistic grows
approximately linearly for $x<(-2\ln\epsilon)^{1/2}\cos\theta$, at which point
it is effectively cut off.  This is the same general behavior as was seen
before for the likelihood ratio $\Lambda(A)$, but there is no longer any
scaling with a prior amplitude.

Another approach is to use a prior distribution $f(A)$ for the amplitude, if
such a distribution is known, to construct an integrated likelihood ratio
$\Lambda=\int\Lambda(A)f(A)dA$; this integrated likelihood ratio would then be
the optimal statistic for detection of all potential signals.  Even if the
prior distribution is not known, it is possible to obtain an approximate
likelihood ratio if $f(A)$ is a slowly-varying function.  One can show that
$\Lambda_{\text{G}}=\int\Lambda_{\text{G}}(A)f(A)dA\approx f(x)\exp(x^2/2)$.
A further approximation neglects the slowly-varying function entirely; one
then obtains $\Lambda_{\text{G}}\approx\exp(x^2/2)$.  Thus,
\begin{equation} \label{e:intLikelihoodRatio}
  \Lambda \approx \frac{e^{x^2/2} + \epsilon e^{(x\sec\theta)^2/2}}
    {1 + \epsilon e^{(x\sec\theta)^2/2}}.
\end{equation}

It is important to generalize the above analysis to the case in which there
are multiple detectors.  Such a situation is deemed to be essential for
gravitational wave searches since a coincident detection is an important
corroboration that a signal is of an astronomical origin rather than internal
to the detector.  As a simple model,%
\footnote{A more general model is being considered by Finn~\cite{finn:1998}.}
consider two gravitational wave detectors with identical but independent noise
(including possible noise bursts) that also have the same response to
gravitational wave signals.  If the noise was purely Gaussian, the likelihood
ratio for the two detectors together would be
$\Lambda_{\text{G}}^{\text{AB}}=
 \Lambda_{\text{G}}^{\text{A}}\times\Lambda_{\text{G}}^{\text{B}}$,
i.e., the product of the likelihood ratios for Gaussian noise in each separate
detector.  The resulting likelihood ratio is thus an monotonically increasing
function of the \emph{sum} of the matched filter values in each detector:
$\ln\Lambda_{\text{G}}^{\text{AB}}\sim x^{\text{A}}+x^{\text{B}}$ where
$x^{\text{A}}=(\uv{u},\op{Q}^{\text{A}}\vv{h}^{\text{A}})$ is the
signal-to-noise ratio in the first detector (A) and
$x^{\text{B}}=(\uv{u},\op{Q}^{\text{B}}\vv{h}^{\text{B}})$ is the
signal-to-noise ratio in the second detector (B).  If the likelihood ratios
are integrated over a slowly-varying prior distribution of possible signal
amplitudes, then one finds 
$\ln\Lambda_{\text{G}}^{\text{AB}}\sim(x^{\text{A}}+x^{\text{B}})^2$,
which also depends on the sum of the two detectors' matched filter values.
(In this case, it is important to integrate over the distribution of
amplitudes \emph{after} multiplying the two detector's likelihood ratios
together since the same signal amplitude should be present in each detector.)
Thus, one would decide that a signal was present if the sum of the matched
filter values in the two detectors exceeded some threshold.  An alternative
strategy would decide that a signal was present only if \emph{each} of the
matched filter values exceeded some threshold, i.e., one thresholds on the
quantity $\min(x^{\text{A}},x^{\text{B}})$ rather than on
$x^{\text{A}}+x^{\text{B}}$.

\begin{figure}[ht]
\begin{center}
\epsfig{file=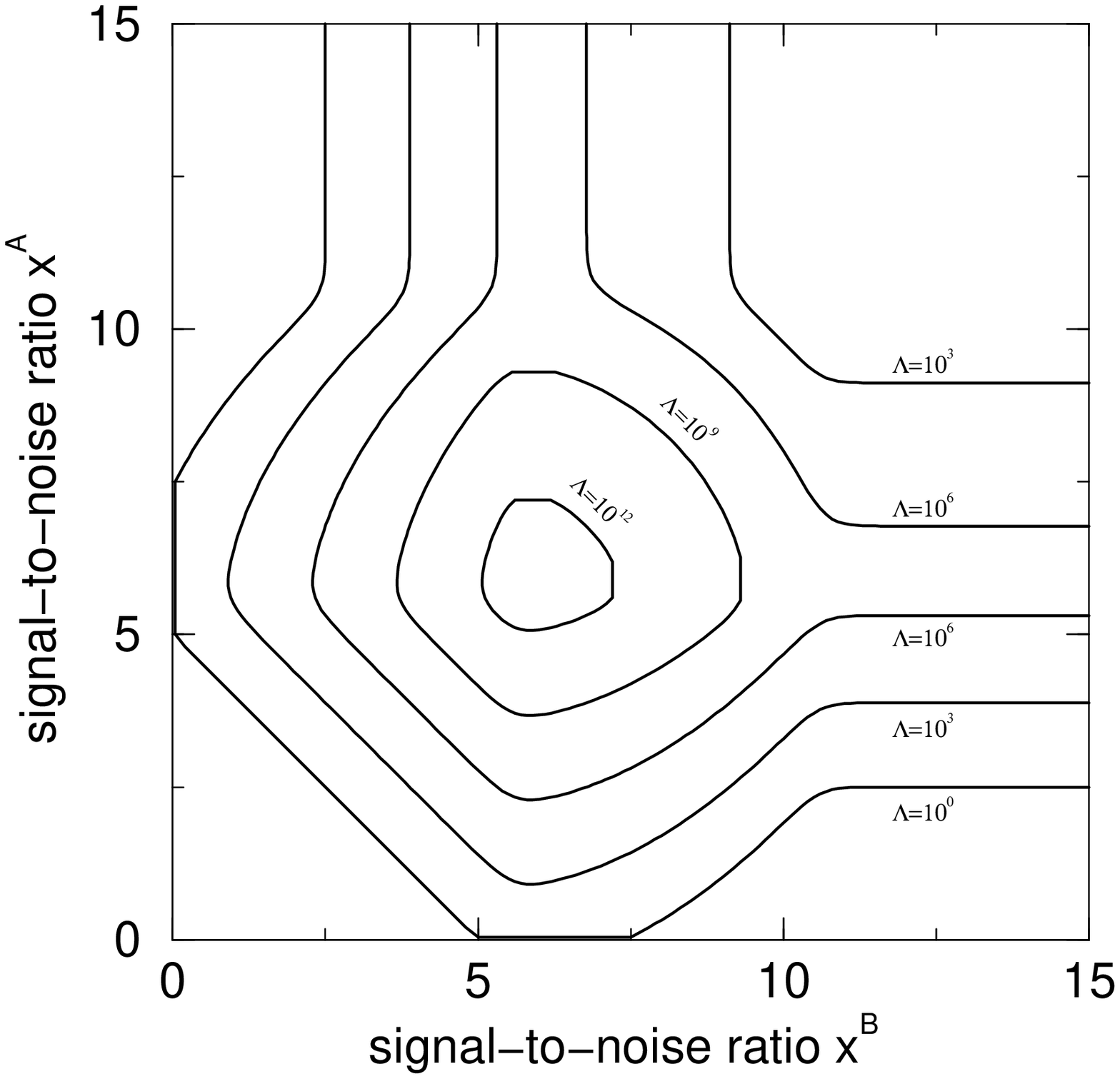,width=0.9\linewidth}\\*
\epsfig{file=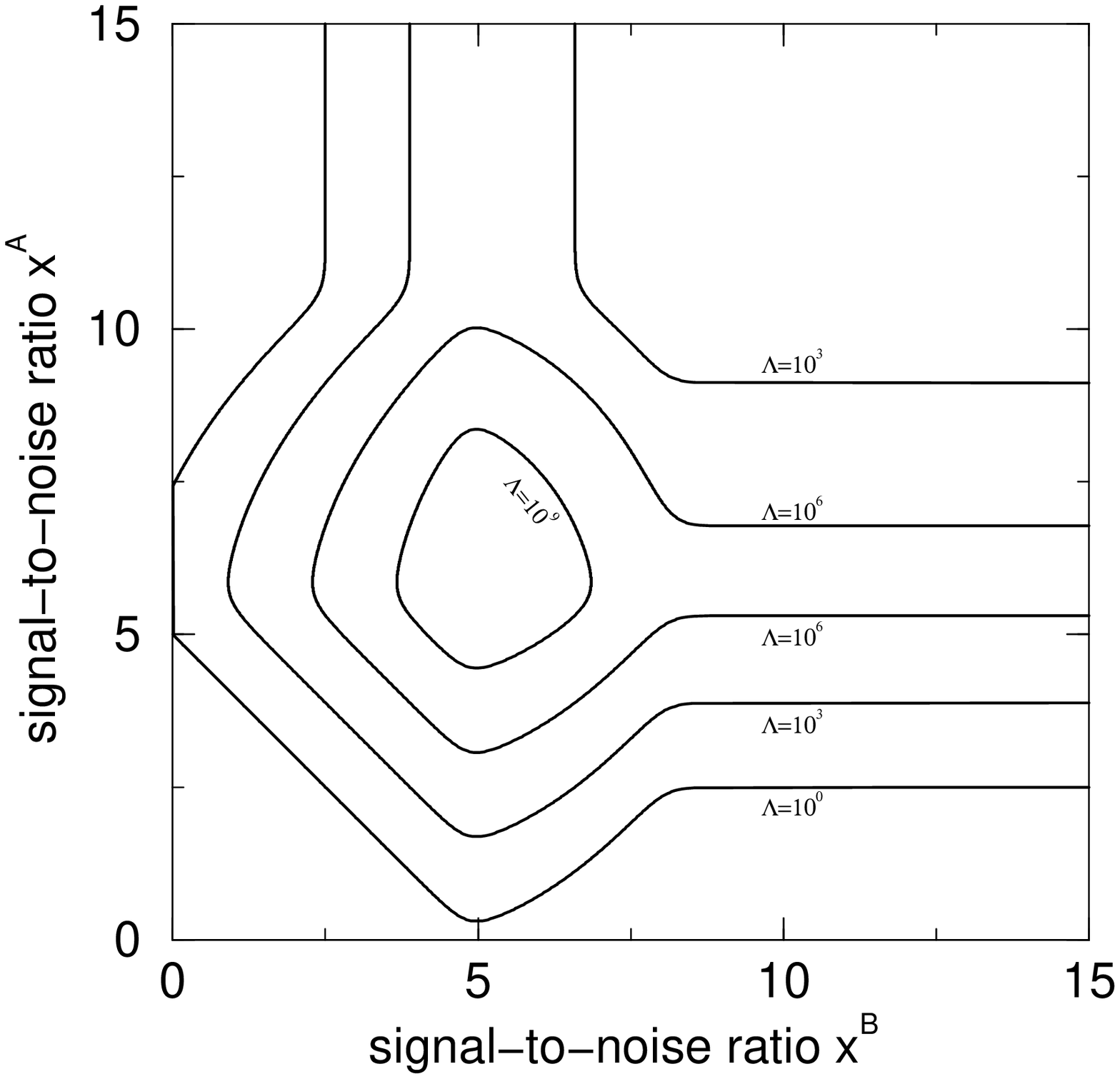,width=0.9\linewidth}
\end{center}
\caption{Contours of constant likelihood ratio for two detectors, A and B,
measuring signal-to-noise ratios $x^{\text{A}}$ and $x^{\text{B}}$.  The
combined likelihood ratio is the product of individual detector likelihood
ratios, which are of the form of Eq.~\ref{e:likelihoodRatio} with
$A=5$ and $\epsilon=2\times10^{-7}$.  The top plot represents the case
when $\theta^{\text{A}}=\theta^{\text{B}}=0$ while the bottom plot represents
the case when $\theta^{\text{A}}=0$ and $\theta^{\text{B}}=\pi/6$.  Notice
that the right side (beyond $x^{\text{B}}\simeq5$) of the bottom plot is 
compressed compared to the top plot, while the left hand side of the two
plots are roughly the same.
\label{f:logLambdaContour}}
\end{figure}

Suppose the likelihood ratios for both detectors,
$\Lambda^{\text{A}}(x^{\text{A}},\theta^{\text{A}})$ and
$\Lambda^{\text{B}}(x^{\text{B}},\theta^{\text{B}})$, are given by
Eq.~(\ref{e:likelihoodRatio}).  Then, the likelihood ratio for the combined
detector is given by the product of $\Lambda^{\text{A}}$ and
$\Lambda^{\text{B}}$.  Figure~\ref{f:logLambdaContour} shows a plot of
contours of constant likelihood ratio as a function of the two measured
signal-to-noise ratios for $\theta^{\text{A}}=\theta^{\text{B}}=0$ and
$\theta^{\text{A}}=0$, $\theta^{\text{B}}=\pi/6$.  At low signal-to-noise
ratios, the contours are approximately lines of constant
$x^{\text{A}}+x^{\text{B}}$---in this regime, the likelihood ratio behaves
like the Gaussian likelihood ratio.  However, for large values of either
$x^{\text{A}}$ or $x^{\text{B}}$, the contours are approximately lines of
constant $\min(x^{\text{A}},x^{\text{B}})$.

The locally optimal two-detector statistic is
$\lambda^{\text{AB}}=\lambda^{\text{A}}+\lambda^{\text{B}}$, where
$\lambda^{\text{A}}$ and $\lambda^{\text{B}}$ both have the form of
Eq.~(\ref{e:localOpt}).  This statistic also depends on $x^{\text{A}}$,
$x^{\text{B}}$, $\theta^{\text{A}}$, and $\theta^{\text{B}}$.  As before, the
contours of constant $\lambda^{\text{AB}}$ for fixed $\theta^{\text{A}}$ and
$\theta^{\text{B}}$ are lines of constant $x^{\text{A}}+x^{\text{B}}$ for low
signal-to-noise ratios but are lines of constant
$\min(x^{\text{A}},x^{\text{B}})$ if the signal-to-noise ratio in one of the
detectors is large.

These results somewhat justify the use of the minimum statistic
$\min(x^{\text{A}},x^{\text{B}})$ when there are two detectors with
independent noise operating.  When one of the signal-to-noise ratios is large
but the other is moderate, the minimum statistic will give approximately the
same value as the locally optimal statistic.  In this case, the effect of the
$\alpha$-based veto in the locally optimal statistic is mimicked in the
minimum statistic.  The locally optimal statistic is better able to deal with
the case when there are noise bursts present in both detectors, but this case
occurs with probability $P_{\text{B}}^2$, so it should be extremely rare.
Moreover, the minimum statistic is less powerful than the locally optimal
statistic for small signal-to-noise ratios in both detectors.  In this case,
the locally optimal statistic is approximately $x^{\text{A}}+x^{\text{B}}$,
which is the same as the optimal statistic for Gaussian noise.  However, the
optimal statistic for Gaussian noise may be unsuitable for the noise model I
have considered because the false alarm probability cannot be reduced below
$\sim P_{\text{B}}$ for reasonable thresholds.  (The minimum statistic suffers
the same problem but around the much lower probability $P_{\text{B}}^2$.)

\begin{figure}[ht]
\epsfig{file=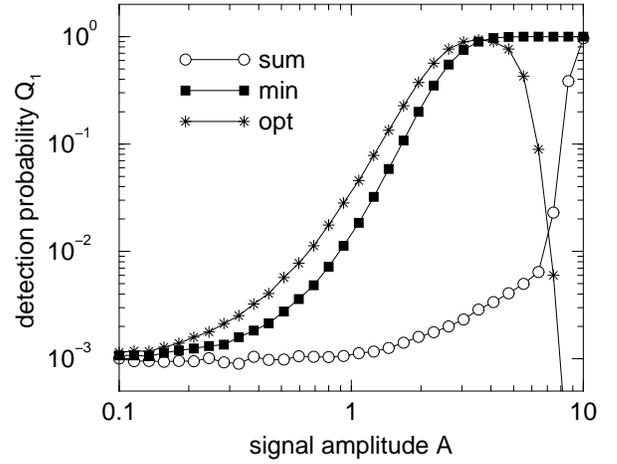,width=0.9\linewidth}
\caption{The operation characteristics, which express the detection
probability $Q_1$ as a function of signal amplitude $A$ for fixed false alarm
probability $Q_0=10^{-3}$ for three multi-detector statistics.  The ``sum''
statistic adds the signal-to-noise ratios of the two detectors; this is the
optimal statistic for purely Gaussian noise.  The ``min'' statistic is
simply the minimum of the two signal-to-noise ratios.  The ``opt''
statistic is the locally optimal statistic for the noise model.  The noise
model has a burst probability of $P_{\text{B}}=1\%$, a maximum burst amplitude
of $R=25$, and a vector space dimension $N=4$.
\label{f:operChar}}
\end{figure}

To see the relative performance of these statistics, it is useful to consider
the operational characteristics of the tests.  For some fixed false alarm
probability $Q_0$, the operational characteristic is the detection probability
$Q_1$ as a function of signal amplitude.  I have computed these curves using
Monte Carlo techniques.  The noise model I used corresponds to that of
Eq.~(\ref{e:noiseBurst}) with burst probability $P_{\text{B}}=1\%$ and maximum
burst amplitude $R=25$.  I fixed the false alarm probability to $Q_0=10^{-3}$
and I examined a vector space dimension of $N=4$.  Figure~\ref{f:operChar}
shows the relative performances of the three statistics in terms of the
detection probability as a function of signal amplitude.  One sees that the
locally optimal statistic performs the best for small amplitude signals (as
expected), but that it has poor performance for large amplitude signals.  This
is because the large amplitude deviations are interpreted as noise bursts and
are suppressed.  (If the locally optimal statistic were used in a real search,
these large events would not be rejected outright but would rather be
subjected to further scrutiny; thus the attenuation of the locally optimal
statistic for large signal amplitudes is somewhat misleading.)  The sum of the
signal-to-noise ratio, which is optimal statistic for Gaussian noise, is seen
to have poor performance.  The reason is that because the false alarm
probability is much smaller than the burst probability, the threshold required
for this statistic becomes unreasonably large.  However, since the false alarm
probability is much higher than the burst probability \emph{squared}, the
minimum statistic performs reasonably well for both small and large amplitude
signals.

The conclusion that can be drawn from this analysis is that search strategies
based on a Gaussian noise model can be significantly improved by considering
a more realistic non-Gaussian noise model, which may contain noise bursts that
occur relatively frequently.  If a method can be devised to detect these noise
bursts and veto data that contains a noise burst, then the use of such a veto
effectively reproduces the locally optimal strategy for the noise model
containing bursts.  However, the necessary veto may not be easily found since
it may not be possible to determine the non-Gaussian properties of the
detector noise to sufficient accuracy.  A viable alternative when two
detectors are operating is to use a coincidence strategy of detection.  Since
the minimum statistic used in the coincidence strategy is a robust statistic,
one would expect that it should have good performance for any realistic noise
model.

I would like to thank Bruce Allen, Patrick Brady, {\'E}anna Flanagan, and Kip
Thorne for their many useful comments on this paper.  This work was supported
by the National Science Foundation grant PHY-9424337.


\end{document}